\documentclass[aps,11pt,onecolumn,prd,preprintnumbers,amsmath,amssymb,nofootinbib,superscriptaddress,showpacs]{revtex4-1}
\pdfoutput=1
\usepackage[T1]{fontenc}
\usepackage{amsthm}
\usepackage{amsfonts}
\usepackage{xcolor}
\usepackage{slashed}
\usepackage{mathtools}
\usepackage{graphicx}
\usepackage{hyperref}
\hypersetup{
    colorlinks,
    citecolor=red,
    filecolor=black,
    linkcolor=blue,
    urlcolor=black
}

\def\ben{\begin{equation}}
\def\een{\end{equation}}

\def\bea{\begin{eqnarray}}
\def\eea{\end{eqnarray}}

\begin{document}

\title{A new energy upper bound for AdS black holes inspired by free field theory}

\author{Krai Cheamsawat}
\email{krai.cheamsawat15@imperial.ac.uk}

\affiliation{Theoretical Physics Group,  Blackett Laboratory, Imperial College, London SW7 2AZ, United Kingdom}

\author{Gary Gibbons}
\email{gwg1@cam.ac.uk}

\affiliation{Department of Applied Mathematics and Theoretical Physics,University of Cambridge, Wilberforce Road, Cambridge CB3 0WA, United Kingdom}

\author{Toby Wiseman}
\email{t.wiseman@imperial.ac.uk}

\affiliation{Theoretical Physics Group, Blackett Laboratory, Imperial College, London SW7 2AZ, United Kingdom}


\begin{abstract}

We consider the toroidally compactified planar AdS-Schwarzschild solution to $4$-dimensional gravity with negative cosmological constant. This has a flat torus conformal boundary metric. We show that if the spatial part of the boundary metric is deformed, keeping it static and the temperature and area fixed, then assuming a static bulk solution exists, its energy is less than that of the AdS-Schwarzschild solution. 
The proof is non-perturbative in the metric deformation. 
While we expect the same holds for the free energy for black hole solutions we are so far are not able to prove it.
In the context of AdS-CFT this implies a $3$-dimensional holographic CFT on a flat spatial torus whose bulk dual is AdS-Schwarzschild has a greater energy than if the spatial geometry is deformed in any way that preserves temperature and area. 
This work was inspired by previous results in free field theory, where scalars and fermions in $3$-dimensions have been shown to energetically disfavour flat space.

\end{abstract} 

\maketitle


\section{Introduction}

Classical matter provides a familiar energetic measure on spatial geometries. For example, given a tension and bending energy for a membrane, one may ask natural questions such as, at fixed area what geometry is energetically preferred by the membrane? The quantum analog of this has recently been studied~\cite{FisWalWis18}. For a field theory at finite temperature defined on an ultrastatic but curved spacetime, $ds^2 = -dt^2 + d\Sigma^2$, for a Riemannian manifold $\Sigma$ with metric $d\Sigma^2 = \bar{g}_{ij}(x) dx^i dx^j$, the energy and free energy in thermal equilibrium are functionals of this spatial geometry. Again it is natural to ask what geometries are energetically preferred. 
We restrict discussion to the case of $3$ spacetime dimensions. Since energy is extensive, we should fix some extensive property of the geometry in making such comparisons, and as in the classical example we choose to compare energies at fixed area.
More precisely we consider a one parameter family of $2$-dimensional spatial geometries $\Sigma(\epsilon)$, with metrics $\bar{g}_{ij}(x; \epsilon)$, which preserve area $A(\epsilon) = \int_\Sigma dx \sqrt{\bar{g}(\epsilon)}$, 
and consider how the free energy $F( \epsilon )$ and energy $E( \epsilon )$ vary with the parameter $\epsilon$ at fixed temperature, $T$. Here the free energy is defined from the equilibrium partition function as $F(T,\epsilon) = - T \ln Z(T,\epsilon)$, and as usual the energy may be computed from this as $E(T,\epsilon) = F(T,\epsilon) - T \partial_T F(T,\epsilon)$. 

We suppose $\epsilon = 0$ is a natural reference geometry to compare energy to -- for example, a homogeneous space such as a round sphere or flat torus. Then we consider the free energy difference $\Delta F(\epsilon) =  F(\epsilon) - F(0)$ as the space is deformed such that $A(\epsilon) = A(0)$, and likewise the energy difference  $\Delta E(\epsilon)$.
Apart from being a natural quantity to fix when comparing energies, preserving area has two additional features~\cite{FisWalWis18,Cheamsawat:2018wkr};
\begin{enumerate}
\item Usually the energetics of field theories have renormalisation counterterm ambiguities, and hence depend on the UV details of the theory. However for the ultrastatic $3$-dimensional case, working at fixed temperature and area and ensuring the only deformation to the theory that is inhomogeneous is due to the spatial metric,  then provided the theory has a diffeomorphism invariant regulator  in fact these energy differences are independent of such ambiguities. We note that does not hold in higher dimensions.
\item
It implies that for a space which is homogeneous (eg. a sphere or torus), for small perturbations of the metric the variation of the free energy is quadratic in the perturbation (rather than linear) and hence may have definite sign.
\end{enumerate}
One may also consider the infinite volume case, such as for deformations of flat space. Then while the area and (free) energy are divergent, for deformations of the spatial geometry with sufficiently quick fall off, the difference in free energy $\Delta F$ from flat space will be finite. The analog of keeping the area fixed is then that the variation of the area functional with respect to $\epsilon$ vanishes.
In~\cite{FisWalWis18} it was shown that the variation of the free energy, $\Delta F(\epsilon)$, of $3$-dimensional relativistic free scalar and fermion fields as a functional of such a perturbation to flat space is always negative to leading quadratic order. This is true for any such temperature and area preserving deformation, for any mass of the fields, and also for any marginal scalar curvature coupling in the scalar field case. It was shown~\cite{Cheamsawat:2018wkr} to be negative for $3$-dimensional holographic CFTs to quadratic order in the perturbation. As we show later, in fact $\Delta E(\epsilon)$ is also negative at quadratic order for these theories. Thus these free fields and holographic CFTs in $2+1$-dimensions appear to energetically disfavour flat space at leading order in perturbations of it.
Since $3$-dimensional free Dirac fermions naturally arise in 2-d crystalline materials, such as monolayer graphene~\cite{Graphene,GrapheneDirac1,Fialkovsky:2016kio,GrapheneDirac}, this energetic decrease for perturbations of flat space may have relevance for the dynamics of these materials, and possibly be related to their tendency to ripple~\cite{FisWalWis18}.
We are then led to the next natural question; 
\begin{itemize}
\item \emph{
What are the global properties of this energetic measure on geometry? } 
\end{itemize}
Is the energy or free energy always bounded above by that for flat space for these theories? Or is it only negative for perturbative deformations of flat space, and for large deformations the energy rises above the flat space value? While in the field theory context, even in free field theory, this question appears to be analytically intractable for general deformations,\footnote{One may hope to address it analytically for specific deformations of homogeneous spaces preserving sufficient symmetry, for example~\cite{Bobev:2017asb}.} we shall see that for $3$-dimensional holographic CFTs then AdS-CFT will allow us to answer this question, at least in the case of energy. 

We now focus on the finite temperature and volume case, where our reference geometry is a flat torus. Then AdS-CFT \cite{Maldacena:1997re,Witten:1998qj,Gubser:1998bc} rephrases the question in terms of the energetics of asymptotically locally AdS gravitational solutions. The behaviour of a $3$-dimensional holographic CFT with large `effective central charge', $c_T = \frac{\ell^2}{16 \pi G}$, where $\ell$ is the AdS length and $G$ the 4-dimensional bulk Newton constant, is given by solutions of a dual bulk gravitational theory.
When the CFT is on a flat spacetime a bulk dual solution is planar AdS-Schwarzschild. Taking the spatial geometry to be a flat torus, $\Sigma(0)$, we toroidally compactify this solution to give what we henceforth call  \emph{toroidal AdS-Schwarzschild}.
Depending on whether there are bulk fermion fields, and what boundary conditions they have about the toroidal cycles, another possible bulk solution is the \emph{AdS-soliton}~\cite{Horowitz:1998ha}. 
Except for the final discussion, we will assume that there are bulk fermion fields and take periodic boundary conditions for them so that this solution is not allowed. Then we expect the bulk dual for thermal equilibrium is toroidal AdS-Schwarzschild.
\footnote{
We do not know of a proof that the bulk spacetime must be toroidal AdS-Schwarzschild under the condition on fermions described here. The most general static exact solutions whose constant radius sections are flat are the planar AdS Kasner family~\cite{Blanco-Pillado:2015dfa,Zofka:2007bi} (with the radial direction playing the role of the usual `Kasner time') which includes AdS-Schwarzschild and the AdS-soliton as its only smooth members. 
However these are not the only solutions. For example, in the case that fermion boundary conditions do admit the soliton, we expect other solutions exist, such as the `plasma-ball' black hole and its generalisations~\cite{Aharony:2005bm}. 
}
We deform the CFT metric to $ds^2 = -dt^2 + d\Sigma^2(\epsilon)$ preserving temperature and area. Provided no other sources are turned on we expect that, at least for reasonably large deformations, static bulk solutions continue to exist within the `universal sector' (see for example \cite{Marolf:2013ioa}), so are  solutions to  pure gravity with a negative cosmological constant.\footnote{In the global AdS context there are claims such bulk solutions do exist for arbitrary deformations of the boundary metric~\cite{Anderson:2002xb} that have positive scalar curvature.  We do not know of a proof in the toroidal context where deformations necessarily result in scalar curvatures that are not everywhere positive.}
Then the CFT question is translated to the gravitational question;
\begin{itemize}
\item 
 \emph{Is the (free) energy of a $4$-dimensional static solution of gravity plus negative cosmological constant with conformal boundary metric $ds^2 = -dt^2 + d\Sigma^2(\epsilon)$ less than that of toroidal AdS-Schwarzschild at the same temperature and boundary area?
 }
\end{itemize}
The energy and free energy here are determined from the renormalised on-shell action, or equivalently from the renormalised boundary stress tensor~\cite{Balasubramanian:1999re,Henningson:1998gx} and the entropy of any horizons. 
For a $3$-dimensional boundary the quantities $\Delta F(\epsilon)$ and $\Delta E(\epsilon)$ have no renormalisation ambiguities.

It is precisely for small perturbations of flat space
 that the above question was answered in~\cite{Cheamsawat:2018wkr} in the affirmative. However here we are interested in the non-perturbative question, assuming a bulk solution with the prescribed conformal boundary metric exists.
We are currently not able to prove such a statement in the affirmative for the free energy, but interestingly can prove it for the energy.
The result we find may be viewed as a `finite temperature' version of a previous result that shows deformations of $4$-dimensional planar AdS~\cite{HicWis15} that are Einstein, and deform the boundary spatial metric away from being the flat torus, decrease the energy.

We can also prove the answer is yes for certain `generalised energies', defined in terms of energy, temperature and entropy, $S$, as $\mathcal{F}_k = E - k T S$, where our proof works for $0 \le k \le \frac{2}{5}$ (but not up to the case of free energy $k=1$).
Energy, or such a generalisation, is obviously not the natural quantity to consider at finite temperature, and we might expect a proof could be made also in the free energy case too, although as we discuss later there would have to be some additional requirement that a bulk horizon exists, and perhaps a restriction on its topology too.

We begin the paper in Section~\ref{sec:review} with a brief review of the $3$-dimensional free field and holographic CFT results, and show they not only imply the leading free energy variation is negative, but also the energy is too. In Section~\ref{sec:bhbound} we collect some useful results and briefly review the argument in~\cite{HicWis15} that at zero temperature solutions of the bulk Einstein equations with conformal boundary that is a deformation of a flat toroidal boundary space lead to negative energy. Then we give our argument that at finite temperature, a bulk solution whose conformal boundary spatial geometry is a deformed torus, has lower energy than that of the toroidal AdS-Schwarzschild solution with the same temperature and boundary area.
We end the paper with a brief summary and discussion.

\section{Previous perturbative results}
\label{sec:review}

Consider a relativistic $(2+1)$-dimensional QFT on a product of time with the Riemannian 2-spaces $\Sigma(\epsilon)$, taking these all to have the same area.
We compute the free energy $F = F(T, \epsilon)$ at temperature $T = 1/\beta$ from the partition function, and its variation is given by,
\begin{eqnarray}
\label{eq:varyDF}
\frac{d F}{d\epsilon} =   \frac{1}{2}  \int d^2x \sqrt{{\bar{g}}} \langle T_{ij} \rangle_{\Sigma(\epsilon)} \frac{d \bar{g}^{ij}}{d\epsilon} 
\end{eqnarray}
where $\langle T_{ij} \rangle = - \frac{2}{\sqrt{{\bar{g}}}} \frac{\delta \ln Z}{\delta {\bar{g}}^{ij}}$ are the spatial components of the stress tensor one-point function in the thermal vacuum. 
This expression derives directly from the path integral and can be thought of as a generalisation of  `$\delta F = p\, \delta V$'.
For a $3$-dimensional theory with diffeomorphism invariant UV regulator, which is only deformed by the metric (ie. all other couplings are constant) we renormalise the UV behaviour of this one-point function with cosmological and Einstein-Hilbert counterterms in the action. However the former does not contribute to the variation above since the family $\Sigma(\epsilon)$ preserves area, and the latter does not, as for our ultrastatic metric the $3$-dimensional Einstein-Hilbert term becomes a two dimensional one, which is topological~\cite{FisWalWis18,Cheamsawat:2018wkr}. Hence the variation $\Delta F(\epsilon) = F(\epsilon) - F(0)$ is insensitive to the finite part of the coefficients of these counterterms, and  is therefore independent of  renormalisation counterterm ambiguities.

Consider a one parameter family such that $\Sigma(0)$ is Euclidean flat space, with $\bar{g}_{ij}(x;0) = \delta_{ij}$. Then since the space is two dimensional we may write a general perturbation of flat space in conformally flat coordinates, so, 
\begin{eqnarray}
\bar{g}_{ij}(x ; \epsilon) = \delta_{ij} \left( 1 + 2 \epsilon f(x)  \right) + O(\epsilon^2) 
\end{eqnarray}
and area preservation in this infinite volume case implies the area functional does not vary with $\epsilon$ so that in particular, $\int d^2x f(x) = 0$. Fourier decomposing the function $f$ as,
\begin{eqnarray}
\label{eq:Fourierf}
f(x)  & = & \int d^2k \, e^{ i k_i x^i} \tilde{f}(k_i) 
\end{eqnarray}
then provided that the only inhomogeneous deformation of the theory is due to the spatial geometry then the leading variation of the free energy will be quadratic in $\tilde{f}(k_i)$ taking the form~\cite{Cheamsawat:2018wkr},
\begin{eqnarray}
\label{eq:quadFvary}
\Delta F(T, \epsilon) & = & - \epsilon^2 \int d^2k \, a( T, k ) \left| \tilde{f}(k_i) \right|^2 + O\left( \epsilon^3 \right)
\end{eqnarray}
for a function $a(T, k)$ depending only on temperature and the magnitude of the wavevector, $k = | k_i |$.
For a free scalar field, mass $m$ and scalar curvature coupling $\xi$, and a Dirac fermion with mass $m$, explicit computation yields~\cite{FisWalWis18};
\begin{eqnarray}
a_{\mathrm{s},\mathrm{f}}(T, k) = - q T k^4 \int_0^\infty d\lambda \, e^{-M^2 \lambda} \Theta_q(T^2 \lambda) I_{\mathrm{s},\mathrm{f}}(k^2 \lambda)
\end{eqnarray}
where we take $q = - \frac{1}{2}$ in the scalar (s) case and $q = +1$ in the fermion (f) case. 
Defining~$\mathcal{F}(\zeta) = \zeta^{-1} e^{-\zeta^2} \int_0^{\zeta} d\zeta' \, e^{(\zeta')^2}$ then,
\begin{eqnarray}
I_\mathrm{s}(\zeta) &=& -\frac{\pi}{4 \zeta^2} \left[6 +  \zeta ( 1 - 8 \xi)  - \left( 6 + 2 \zeta (1 - 4 \xi) + \frac{\zeta^2}{2}  (1 - 4 \xi)^2 \right) \mathcal{F}\left( \frac{\sqrt{\zeta}}{ 2} \right) \right]
\\
I_\mathrm{f}(\zeta) &=& \frac{\pi}{4 \zeta^2} \left[\left(6 + \zeta\right)\mathcal{F}\left(\frac{\sqrt{\zeta}}{ 2} \right) - 6 \right] 
\end{eqnarray}
and where,
\begin{eqnarray}
\Theta_q(\zeta) = \sum_{n=-\infty}^{\infty} e^{- (2 \pi)^2 (n-q+1/2)^2 \zeta} \; .
\end{eqnarray}
 In both cases we see explicitly that $a(T, k)$ is positive since both $\Theta_{q} (T^2 \lambda) \ge 0$ and $I_{f,s}(k^2 \lambda) \ge 0$. This results in a negative quadratic free energy variation. Likewise in~\cite{Cheamsawat:2018wkr} the function $a(T, k)$ was determined for a $3$-dimensional holographic CFT. Perturbations of the dual AdS-Schwarzschild solution that deform the boundary space, and the resulting response of the stress tensor one-point function (the `geometric polarization'~\cite{Emparan:2017qxd}) were computed, and from these $a(T, k)$ was extracted. This was again found to be positive (and strikingly similar in form to that of the free Dirac fermion CFT).

\begin{figure}
\centering
\includegraphics[width=0.8\textwidth,page=1]{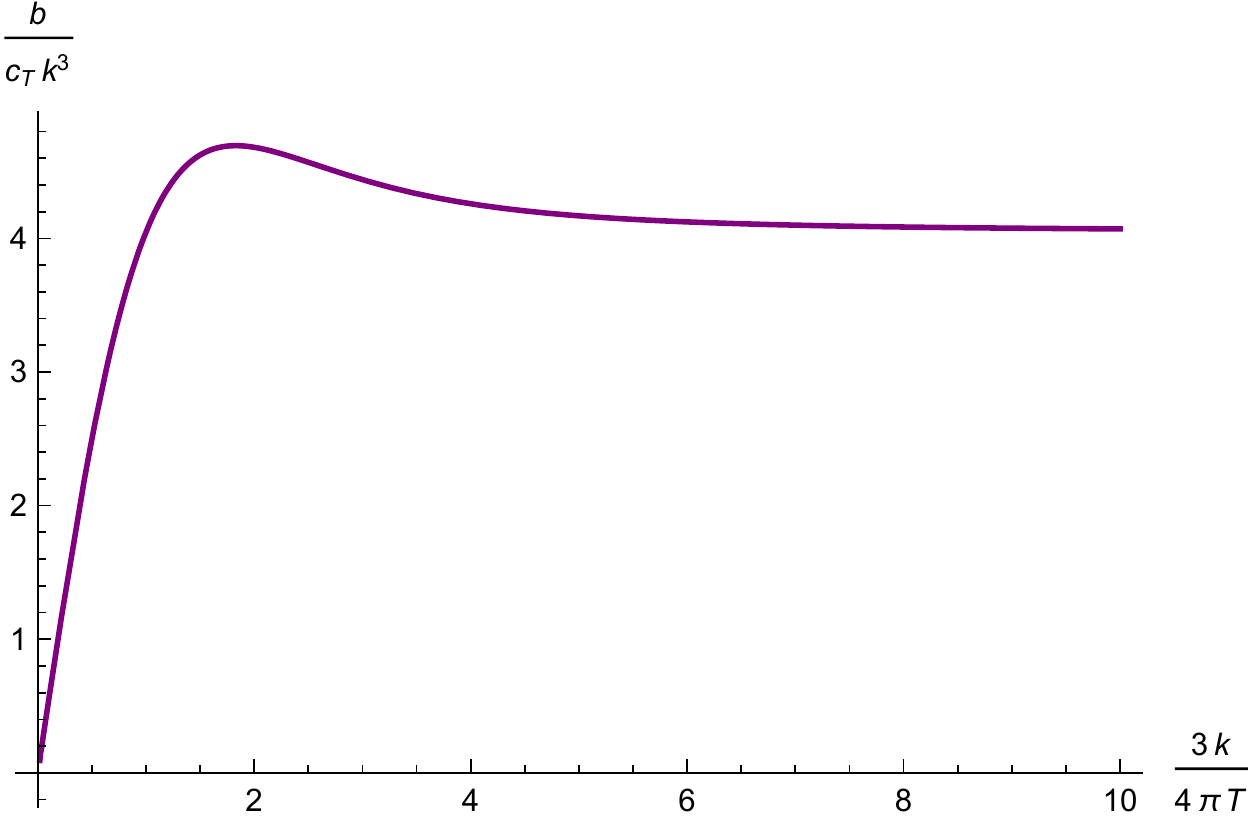}
\caption{Graph of the function $b(T, k)$ normalised by $k^3$ and the CFT central charge, $c_T$, against $3 k / (4 \pi T)$ for the holographic CFT. Its positivity implies that $\Delta E(\epsilon)$ is negative to leading quadratic order for perturbative deformations of flat space, for any perturbation and temperature.
}
\label{fig:bcurve}
\end{figure}

Of relevance for the results in this paper, instead of considering the variation of free energy we may consider the variation of energy $\Delta E(\epsilon)$. Then using $\Delta E = \Delta F - T \partial_T \Delta F |_\epsilon$, we have,
\begin{eqnarray}
\Delta E(T,\epsilon) &=& - \epsilon^2 \int d^2k \, b( T, k )  \left| \tilde{f}(k_i) \right|^2 + O\left( \epsilon^3 \right) \; , \quad b( T, k ) = a( T, k ) -  T \frac{\partial}{\partial T} a( T, k )
\end{eqnarray}
and one can quickly see that in the free scalar and fermion cases $b(T,k) \ge 0$ over the domain of interest. In the holographic case, using the numerical solution of $a(T, k)$ we may compute $b(T,k)$ which is shown in figure~\ref{fig:bcurve} and is indeed again positive.
Thus in these theories flat space is energetically disfavoured perturbatively for both the free energy, and the energy.

While the holographic case was discussed in~\cite{Cheamsawat:2018wkr} in the infinite volume setting, by restricting to suitably periodic deformations of the metric one can  toroidally compactify and obtain the same result (again assuming fermion boundary conditions so that the dual to the flat torus is AdS-Schwarzschild rather than the soliton). Thus we learn that in the holographic case, at fixed temperature, for arbitrary perturbations of the boundary flat torus leaving its area fixed, both $\Delta F$ and $\Delta E$ vary quadratically in the amplitude of the perturbation in the negative direction as the bulk dual is perturbed away from AdS-Schwarzschild. 
The purpose of the remainder of the paper is to show that the flat torus is energetically disfavoured non-linearly in the perturbation.

\section{Black hole bound}
\label{sec:bhbound}

Consider a static $4$-dimensional solution to pure gravity with a negative cosmological constant, $R^{(4)}_{AB} = - \frac{3}{\ell^2} g^{(4)}_{AB}$. We may write the spacetime metric in terms of the Riemannian optical geometry $\mathcal{M}$ (see for example~\cite{HicWis15}), with optical metric $d\mathcal{M}^2 = g_{ab}(x) dx^a dx^b$ as,
\begin{eqnarray}
ds^2_{(4)} = \frac{\ell^2}{Z^2(x)} \left( - dt^2 + g_{ab}(x) dx^a dx^b \right) \, .
\end{eqnarray}
We may decompose the Einstein equations over $\mathcal{M}$ as,
\begin{eqnarray}
\nabla^2 \left( \frac{1}{Z^2} \right) = \frac{6}{Z^4} \; , \qquad R_{ab} = - \frac{2}{Z} \nabla_a \partial_b Z
\end{eqnarray}
where $R_{ab}$ is the Ricci tensor of the optical metric, and all expressions are written covariantly with respect to $\mathcal{M}$. 
The AdS length scale $\ell$ is scaled out of these equations, which differ from those without negative cosmological constant only in the presence of the term on the righthand side of the first equation. 
We will be interested in the optical Ricci scalar, $R$, which obeys $R = \frac{6}{Z^2} \left( 1 - (\partial Z)^2 \right)$. 

We now consider solutions with a toroidal topology spatial conformal boundary metric. 
We require these solutions to be smooth, and we allow the spacetime to end on a Killing horizon with a finite number of components, $N$, all with the same Hawking temperature $T$, or to have no horizon (although we expect this is disallowed by the fermion boundary conditions).
This implies $\mathcal{M}$ has a boundary, $\partial \mathcal{M}_\infty$, associated to the spacetime conformal boundary and also may have asymptotic ends $\partial \mathcal{M}_{H_k}$ for each of the $N$ Killing horizon components, $H_k$, so $k=1,2,\ldots, N$, but has no other boundaries or asymptotic regions. Then we may write,
$\partial \mathcal{M} = \partial \mathcal{M}_\infty \cup \partial \mathcal{M}_{H_1} \cup \partial \mathcal{M}_{H_2} \cup \ldots \cup \partial \mathcal{M}_{H_N}$. 
For small deformations of the boundary torus away from being flat, we expect solutions to exist that are small deformations of AdS-Schwarzschild, and these will have a single toroidal horizon.
We do not know if solutions exist with a different number of horizon components to AdS-Schwarzschild. 
There are examples for fermion boundary conditions that allow for the AdS-soliton. Then the soliton itself has no horizon, and one may imagine exotic solutions with multiple horizon components, constructed by taking toroidally compactified `plasmaball' solutions~\cite{Aharony:2005bm,Figueras:2014lka}, so an AdS-soliton with one or more `small' black holes contained within it. However these examples and their generalisation to non-flat boundary tori are presumably not allowed due to our fermion boundary conditions.

\subsection{Asymptotic and horizon behaviours}
\label{sec:boundaries}

The spacetime conformal boundary is at $Z(x) = 0$ and from the perspective of the optical geometry it results in an actual boundary $\partial \mathcal{M}_\infty$ of $\mathcal{M}$. We wish the spacetime conformal boundary to have conformal class given by  $ds^2 = - dt^2 + d\Sigma(\epsilon)^2$, and we may think of this as a `Dirichlet' boundary condition for the bulk equations.
Choosing the natural conformal frame defined by $Z$, where we take boundary time to be the extension of the bulk time, then the metric induced from $d\mathcal{M}^2 = g_{ab}(x) dx^a dx^b$ on the boundary $\partial \mathcal{M}_\infty$ is then that of $\Sigma(\epsilon)$. 
In this conformal frame, computing the time-time component of the renormalised boundary stress tensor will yield the energy density $\rho(x^i)$ of the spacetime defined relative to the bulk Killing vector field $\partial /\partial t$.  Explicitly near this boundary the geometry has a Fefferman-Graham expansion, where taking $x^a = (z, x^i)$ one finds
using~\cite{Skenderis:2002wp,deHaro:2000vlm} that,
\begin{eqnarray}
ds^2_{(4)} = \frac{\ell^2}{z^2} \left( dz^2  - \left(1 + \frac{1}{4} \bar{R}(x)  z^2 - \frac{1}{3 c_T} \rho(x) z^3 + \ldots \right) dt^2 + \left( \bar{g}_{ij}(x) + \ldots \right) dx^i dx^j \right) 
\end{eqnarray}
so that the conformal boundary in the frame defined by $z$ is,
\begin{eqnarray}
ds^2_{\infty} = - dt^2 + \bar{g}_{ij}(x)  dx^i dx^j 
\end{eqnarray}
as required, and $\bar{R}$ is the Ricci scalar of the metric $\bar{g}_{ij}$.
We may compute that $Z$ and the optical Ricci scalar have expansions~\cite{HicWis15},
\begin{eqnarray}
Z(z,x) & = & z - \frac{1}{8} \bar{R}(x) z^3 + \frac{1}{6 c_T} \rho(x) z^4 + \ldots \nonumber \\ 
R(z,x) & = & 3 \bar{R}(x) - \frac{6}{c_T} \rho(x) z + \ldots\, .
\end{eqnarray}
The total energy $E$ of the spacetime is given by integrating the energy density $\rho$ over the boundary metric $\Sigma(\epsilon)$, and we will also be interested in the boundary area $A(\Sigma)$ and Euler characteristic;
\begin{eqnarray}
E(\Sigma) & = & \int_\Sigma d^2x \sqrt{\bar{g}} \rho  \; , \quad \chi(\Sigma)  = \frac{1}{4\pi} \int_\Sigma d^2x \sqrt{\bar{g}} \bar{R} \, .
\end{eqnarray}

Now consider a Killing horizon component, $H_k$, with non-zero surface gravity $\kappa$. This results in an asymptotic end, $\partial \mathcal{M}_{H_k}$, in the optical geometry. As we are considering thermal equilibrium, all such components will have the same surface gravity, which is determined by the temperature as $T = \frac{\kappa}{2\pi}$.  We may write the metric using a normal coordinate $r$ and horizon coordinates $y^i$, so that the horizon is at $r = 0$ and then near it the metric behaves as,
\begin{eqnarray}
ds^2_{(4)} =  dr^2  - \kappa^2 r^2 \left( 1 + O(r^2) \right) dt^2 + \left( h_{ij}(y) + O(r^2) \right) dy^i dy^j 
\end{eqnarray}
with $h_{ij}$ the horizon metric. This behaviour yields~\cite{HicWis15},
\begin{eqnarray}
\frac{1}{\ell} Z(r, y) & = & \frac{1}{\kappa\, r} + \frac{1}{12 \kappa} R^{(h)}(y) r + \ldots \nonumber \\ 
R(r, y)  & = & -6 \kappa^2 + 3 \kappa^2 \left( \frac{2}{\ell^2} + R^{(h)}(y) \right) r^2 + \ldots
\end{eqnarray}
with $R^{(h)}$ the Ricci scalar of the horizon metric $h_{ij}$. As usual the entropy of this horizon component is given by $S_{H_i} = \frac{4 \pi  c_T}{\ell^2} A(H_k)$ with $A(H_k) = \int_{H_k} d^2y \sqrt{h}$ giving its area. Its Euler characteristic is $\chi(H_k) = \frac{1}{4\pi} \int_{H_k} d^2y \sqrt{h} R^{(h)}$.

\subsection{Toroidal AdS and AdS-Schwarzschild}

For Poincare (or planar) AdS the optical geometry is flat space, $d\mathcal{M}^2 = \delta_{ij} dx^i dx^j + dz^2$, so that the optical Ricci scalar vanishes, $R = 0$, and $Z = z$ is linear in the natural radial coordinate. This solution ends at an extremal horizon as $z \to \infty$. 
At finite temperature this solution becomes planar AdS-Schwarzschild,
\begin{eqnarray}
\label{eq:Sch}
ds^2_{(4)} = \frac{\ell^2}{z^2} \left( - \left( 1 - \left( \frac{z}{z_0} \right)^3 \right) dt^2 + \delta_{ij} dx^i dx^j + \left( 1 - \left( \frac{z}{z_0} \right)^3 \right)^{-1}  dz^2 \right)
\end{eqnarray}
which has a non-extremal horizon at $z = z_0$. Then the energy density and surface gravity are,
\begin{eqnarray}
\frac{1}{c_T} \rho = \frac{2}{z_0^3} \; , \quad \kappa = \frac{3}{2z_0}
\end{eqnarray}
and for this solution $Z$ and the optical Ricci scalar, $R$, go as,
\begin{eqnarray}
\label{eq:RZreln}
Z & = & \frac{z}{\sqrt{1 - \left( \frac{z}{z_0} \right)^3}} \; , \quad R = - \frac{3}{2 z_0^2} \left( \frac{z}{z_0} \right) \left( 8  + \left( \frac{z}{z_0} \right)^3 \right)
\end{eqnarray}
and this defines a relation $R = R(Z)$ (which is invertible over the domain of interest, $Z \ge 0$) between these two scalar quantities of the optical geometry for this solution which will be important in what follows.

We wish to focus here on solutions with finite boundary area, and take the boundary geometry to be a deformation of the flat torus.
For both the AdS and AdS-Schwarzschild solutions we see the boundary metric is flat space, $ds^2 = - dt^2 + dx^i dx^i$ and is compactified to a torus by taking $x^i \sim x^i + L$. The same identification acts on the bulk spacetime, and for the finite temperature case gives toroidal AdS-Schwarzschild with a Killing horizon that has toroidal geometry. We will not consider the zero temperature compactified AdS case further, although we note that this identification acts on the extremal horizon of Poincare AdS to give a singularity.\footnote{
This may be seen by embedding (unit radius) Poincare AdS as the quadric $U V - X^\mu X_\mu = 1$ taking,
$U = \frac{1}{z}$, $V = z + \frac{1}{z} \eta_{\mu\nu} x^\mu x^\nu$, $X^\mu = \frac{1}{z} x^\mu$,
where $\mu = 0,1,2$, $x^\mu = (t, x, y)$ and $\eta_{\mu\nu} = \mathrm{diag}(-1,1,1)$. 
The induced metric is then of our form above in equation~\eqref{eq:Sch} with $z_0 = 0$ and $\ell = 1$. The extremal horizon is the null hyperplane $U = 0$, and Poincare AdS is the portion $U \ge 0$ of the quadric.
Consider the translation $x \to x + L$ which is an element of the AdS $SO(3,2)$ isometry. This yields $V \to V + 2 L X^1 + U L^2$, $X^1 \to X^1 + U L$
with $U$ and the other $X^\mu$ invariant. This acts freely except for the curve  $X^1 = 0$ on the extremal horizon $U = 0$. Hence identifying with $x \sim x + L$ leads to an orbifold singularity on the extremal horizon. 
}

\subsection{Zero temperature energy bound}

We may now briefly review a related previous result~\cite{HicWis15,FisWis17} in our context. We begin considering finite temperature $T > 0$ and take $\Sigma(\epsilon)$ to be deformations of a flat torus $\Sigma(0)$. As discussed above we assume the only bulk solution for flat boundary torus is toroidal AdS-Schwarzschild. As we deform in $\epsilon$ to a non-flat boundary $ds^2 = -dt^2 + d\Sigma(\epsilon)^2$, we assume a smooth static bulk solution exists with the correct boundary geometry and ending on a toroidal Killing horizons with appropriate surface gravity fixed by the temperature.
Such a static bulk solution obeys the neat relation,
\begin{eqnarray}
\label{eq:zeroTbound}
\nabla^2 R = - 3  \tilde{R}_{ab} \tilde{R}^{ab} \le 0
\end{eqnarray}
where $\tilde{R}_{ab} = R_{ab} - \frac{1}{3} g_{ab} R$ is the trace removed optical Ricci tensor.

This identity has an interesting history. It is found (written in ADM rather than optical form) in the work of Robinson~\cite{Robinson} and Lindblom~\cite{Lindblom} without cosmological constant and was extended to AdS space in~\cite{Boucher:1983cv}. It also appeared earlier in the work of Buchdahl~\cite{Buchdahl}, which in turn followed from a result of Lanczos~\cite{Lanczos:1938sf} which is closely related to the Gauss-Bonnet theorem, although at the time the authors of~\cite{Boucher:1983cv} were unaware of that work.
Buchdahl writes the righthand side of the identity above as the square of the 4-tensor, 
\begin{eqnarray}
^{(4)}J_{\alpha \beta \gamma \delta} = {^{(4)}R_{\alpha \beta \gamma  \delta}}-
\frac{\Lambda}{3}  \Bigl( {^{(4)}g_{\alpha \gamma}} \,{ ^{(4)}g_{\beta \delta}}  -
{^{(4)}g_{\alpha \delta}} \, {^{(4)}g_{\beta \gamma}}       \Bigr ) 
\end{eqnarray}
where we note that $\frac{2}{\ell^4} Z^4 | \tilde{R}_{ab} |^2 =  | J^{(4)}_{\alpha\beta\mu\nu} |^2$. This quantity vanishes if and only if the metric $^{(4)}g_{\mu \nu}$
is of constant curvature and
hence, locally at least, Anti-de-Sitter spacetime. Thus only for such a spacetime is the inequality $\nabla^2 R  \le 0$ saturated.

Now assuming the existence of a smooth, static but not everywhere planar
solution to the Einstein equations, we may
integrate this relation over the optical geometry $\mathcal{M}$ and using the divergence theorem we find contributions from the components of $\partial\mathcal{M}$, the boundary $\partial\mathcal{M}_\infty$ and any horizons, $\partial\mathcal{M}_{H_k}$, so,
\begin{eqnarray}
\int_{\partial \mathcal{M}_\infty} dA^a \partial_a R + \sum_{k=1}^N \int_{\partial \mathcal{M}_{H_k}} dA^a \partial_a R  \le 0
\end{eqnarray}
where $dA^a$ is the appropriate outward directed area element. Evaluating these using the results in section~\ref{sec:boundaries} yields,
\begin{eqnarray}
\label{eq:bound1}
\frac{1}{c_T} \left( E - T S \right) \le 8 \pi^2 T \chi(H) 
\end{eqnarray}
where $S = \sum_{k=1}^N S_{H_k}$ is the total entropy (of all horizon components) and $\chi(H) = \sum_{k=1}^N \chi(H_k)$ is the Euler characteristic of the union of all the bulk horizon components.
Considering the zero temperature limit, $T \to 0$, then provided the entropy is bounded, as we should expect, then we learn that $E \le 0$, and hence the energy is non-positive in this zero temperature limit. Since (given our assumed fermion boundary conditions) for the flat torus boundary the only dual solution is toroidal AdS-Schwarzschild, which limits to toroidal AdS at zero temperature, with zero energy, so we see in this case that $\Delta E \le 0$ at zero temperature. Note that this is a non-perturbative statement in the deformation.

It is not obvious that the surface terms above evaluate to finite quantities at the conformal boundary, and in bulk dimensions more than 4 this is not the case. It is for this reason that we restrict the discussion in this paper to a 4-dimensional bulk~\cite{HicWis15}. 

Does the bound give a useful result at finite temperature? Suppose the horizon has a single component which is toroidal, as for AdS-Schwarzschild. Then $\chi(H) = 0$ and so we learn that $F \le 0$, so the free energy is non-positive. However since the finite temperature free energy of toroidal AdS-Schwarzschild is already negative, so $F(0) \le 0$, then this does not yield a bound for $\Delta F$.

\subsection{Finite temperature bound}

As we shall now see, an interesting finite temperature bound on energetic differences will require somewhat more sophistication. As above we consider deforming the boundary metric $\Sigma(\epsilon)$ away from a flat torus $\Sigma(0)$,  where the only (given our fermion boundary conditions) bulk solution is toroidal AdS-Schwarzschild,  preserving both temperature, $T$, and boundary area. While we would hope to find a bound on $\Delta F(\epsilon)$ in this case, here we will obtain a bound on the energy difference $\Delta E(\epsilon)$ instead, 
and also on certain `generalised energies' defined as,
\begin{eqnarray}
\mathcal{F}_k =  E - k \, T S 
\end{eqnarray}
so that energy and free energy are the cases $k=0$ and $1$ respectively, and the generalised energy difference is defined as $\Delta \mathcal{F}_k = \mathcal{F}_k(\epsilon) - \mathcal{F}_k(0)$.
 Indeed we obtain the finite temperature result,
\begin{eqnarray}
\label{eq:result}
\boxed{ \frac{1}{c_T} \Delta \mathcal{F}_k(\epsilon) \le 0 }
\end{eqnarray}
for any $0 \le k \le \frac{2}{5}$ under the assumption that a smooth bulk solution exists.
We will make use of the techniques that Robinson used in proving uniqueness of asymptotically flat Schwarzschild~\cite{Robinson}. The idea that the  boundary spatial deformations $\Sigma(\epsilon)$ preserve area are a key part of this bound, and very much inspired by the free field energy bounds discussed above.

From our perspective, the fundamental issue with the bound in equation~\eqref{eq:bound1} above is that it is not saturated by AdS-Schwarzschild, but only by AdS itself. Hence we  seek a similar identity, where a divergence of a quantity that integrates up to surface terms is related to terms of definite sign, and which vanish for toroidal AdS-Schwarzschild. In the asymptotically flat context, such a relation formed the backbone of Robinson's uniqueness argument. It is less clear our result will be useful in proving uniqueness of AdS-Schwarzschild on its own. In a sense it is quite the opposite. Instead of telling us AdS-Schwarzschild is a unique bulk solution, rather it tells us that the bulk solution is not AdS-Schwarzschild when the boundary metric is non-flat. This freedom to deform the boundary metric changes the nature of the uniqueness problem, and any proof of uniqueness for AdS-Schwarzschild (such as~\cite{Chrusciel:2000az} in the higher genus boundary cases) must specify the boundary metric to have an appropriate fixed geometry. 

Two important components of Robinson's relation are that the spatial geometry of Schwarzschild is conformally flat, and that it is a cohomogeneity one metric. 
The same, of course, is true for toroidal AdS-Schwarzschild, and following Robinson we begin by defining the conformally invariant  Cotton tensor $R_{abc}$ of the optical geometry;
\begin{eqnarray}
R_{abc} = \nabla_c R_{ab} - \nabla_b R_{ac} + \frac{1}{4} \left( g_{ac} \partial_b R - g_{ab} \partial_c R \right) \, .
\end{eqnarray}
This vanishes for a conformally flat optical geometry, such as that of toroidal AdS-Schwarzschild. 
Using the bulk equations of motion, which in particular imply $\partial_a R = \frac{6}{Z} \tilde{R}_{ab} \partial^b Z$,  and $\partial_a \left( Z^2 R \right) = 6 Z R_{ab} \partial^b Z$, we find the square of this, which is positive, is,
\begin{eqnarray}
\label{eq:RobinsonAdS}
0 \le R_{abc} R^{abc} & = & \frac{ 2 (\partial Z)^2 }{ Z^2}  \tilde{R}_{ab} \tilde{R}^{ab} - \frac{1}{12} (\partial_a R)^2 \, .
\end{eqnarray}
The consequence of toroidal AdS-Schwarzschild being a cohomogeneity one metric that we utilise is that it has a definite relation $R = R(Z)$ given by equation~\eqref{eq:RZreln}. Let us define a scalar function, $\Phi$, on the optical geometry,
\begin{eqnarray}
\label{eq:phi}
\Phi(x) = \left( R - \frac{6}{Z^2} \right) - H(Z)
\end{eqnarray}
where $H(Z)$ is the function on $Z \ge 0$ chosen so that $\Phi$ vanishes for the toroidal AdS-Schwarzschild with temperature $T$. This is expressed most conveniently using the Schwarzschild coordinate $z$ introduced in~\eqref{eq:Sch}, so $Z(z) = z/\sqrt{1-z^3/z_0^3}$ with $T = 3/(4 \pi z_0)$ and then,
\begin{eqnarray}
H(Z(z)) & = & - \frac{3}{2 z^2} \left(2 + \left( \frac{z}{z_0} \right)^3 \right)^2 \, .
\end{eqnarray}
From this we define a covector field $V_a$,
\begin{eqnarray}
V_a = \partial_a \Phi + \frac{1}{H(Z)} \left( \frac{12}{Z^3} - H'(Z) \right) \Phi \, \partial_a Z 
\end{eqnarray}
which again vanishes for toroidal AdS-Schwarzschild, with temperature $T$, since then $\Phi = 0$ and $\partial_a \Phi = 0$.
Then we take the function $H(Z)$, and from it derive 3 functions $A(Z)$, $B(Z)$, $C(Z)$ by solving the following 
system of differential equations;
\begin{eqnarray}
\label{eq:ode}
A'(Z) & = & \frac{3}{2} \frac{A(Z) }{H(Z) } \left( \frac{12}{Z^3} - H'(Z)  \right)  - 9 B(Z)  \nonumber \\
B'(Z)  & = & \frac{1}{12}  \frac{A(Z) }{H(Z)^2} \left( \frac{12}{Z^3} - H'(Z)  \right)^2  - \frac{3}{Z} B(Z) + \frac{18}{Z^3} \frac{C(Z) }{H(Z)^2} \nonumber \\
C'(Z)  & = & - \frac{3}{Z^3 H(Z) } \left( 10 B(Z) H(Z)  + 12 C(Z) + Z^2 C(Z)  H(Z) \right)
\end{eqnarray}
and using these we define a covector field,
\begin{eqnarray}
\label{eq:J}
J_a(x) = \frac{A(Z)}{9} \partial_a R - 6 B(Z)  \frac{(\partial Z)^2}{Z^2}  \, \partial_a Z + C(Z) \partial_a Z
\end{eqnarray}
whose divergence, upon using the bulk Einstein equations together with these ODEs constraining $A,B$ and $C$, can then be written as,\footnote{
One might be concerned that $(\partial_a Z)^2$ appears in the denominator of the first term on the right-hand side. However the term is regular when this vanishes as we see from equation~\eqref{eq:RobinsonAdS} and the fact that $\partial_a R$ vanishes when $\partial_a Z$ does since $\partial_a R = \frac{6}{Z} \tilde{R}_{ab} \partial^b Z$.
}
\begin{eqnarray}
\label{eq:Robinson}
 \nabla^a J_a   & = & - A(Z) \frac{Z^2}{(\partial Z)^2} \left( \frac{1}{6} R_{abc} R^{abc} + \frac{1}{72} V_a  V^a \right)  -  C(Z) \frac{3}{Z H(Z)^2} \Phi^2 \, .
\end{eqnarray}
The purpose of introducing the various functions is twofold. 
Firstly if we can choose solutions of the ODE system so $A(Z), C(Z) \ge 0$ for all $Z \ge 0$ then the terms on the right-hand side have definite sign for any smooth bulk solution\footnote{Note that since the system is linear in $A, B$ and $C$, we may choose the sign to be positive without loss of generality.}, and this yields a bound $\nabla^a J_a \le 0$ which can be integrated over $\mathcal{M}$ to give a bound on surface terms,
\begin{eqnarray}
\label{eq:RobinsonBound}
\int_{\partial \mathcal{M}}  dA^a J_a   & \le  & 0 \quad \mathrm{if} \quad A(Z), C(Z) \ge 0 
\end{eqnarray}
where $dA^a$ is the outer directed area element. 
Secondly, it tailors the divergence $\nabla^a J_a$ to vanish for toroidal AdS-Schwarzschild with temperature $T$, since then $R_{abc} = 0$ and the function $\Phi$ and covector $V_a$ vanish by their construction. As a consequence the above bound on the surface terms is saturated for toroidal AdS-Schwarzschild with temperature $T$.

Let us now consider the solution to this ODE system, and the resulting surface terms from integrating $\nabla^a J_a$ over $\mathcal{M}$. Using the Schwarzschild coordinate $z$ introduced in~\eqref{eq:Sch}, and defining the dimensionless $\mu = z/z_0$, the general solution to the ODE system for $A$, $B$ and $C$ above is,
\begin{eqnarray}
A(Z(\mu)) & = & \frac{1}{5} \frac{1}{\left( 2+\mu^3 \right)^2} \left( 20 a - 45 b \mu \left(4 - \mu^3 \right) - 6 c \mu^3  \right) \nonumber \\
B(Z(\mu)) & = & \frac{4}{5 z_0} \frac{ \left(1-\mu^3\right)^{3/2}}{\left(2+\mu^3\right)^3} \left(  10 b \left(1 - \mu^3\right) + c \mu^2  \right) \nonumber \\
C(Z(\mu)) & = & \frac{2}{15 z_0^{3}} \frac{\left(1-\mu^3\right)^{3/2}}{\left(2+\mu^3\right)^2} \left( -20 a \mu^3  + 90 b  \left(\frac{2}{\mu^2}   - 5 \mu \right) + 3 c \left(10 + \mu^3\right) \right) 
\end{eqnarray}
for (dimensionless) integration constants $a$, $b$ and $c$. Using these we may evaluate the surface terms due to the spacetime conformal boundary, arising in the optical geometry from $\partial \mathcal{M}_\infty$, and the $k$-th spacetime horizon component, corresponding to $\partial \mathcal{M}_{H_k}$. Since we are considering the temperature $T$ to be fixed, we take all horizon components to have surface gravity corresponding to this temperature. We then obtain,
\begin{eqnarray}
\int_{\partial \mathcal{M}_\infty}  dA^a J_a & = &  \frac{2}{3} a \frac{E}{c_T} - 16 \pi^2 b \,T \chi(\Sigma) - \frac{2}{5} \left( \frac{4 \pi T}{3}\right)^3 c A(\Sigma) \\
\label{eq:horizonterms}
 \int_{\partial \mathcal{M}_{H_k}}  dA^a J_a & = & - \frac{2}{3}  a \frac{T S_{H_k}}{c_T}  \left( \frac{2}{3} - \frac{c}{5 a}  \right) - \frac{(4 \pi)^2}{3} T \, \chi(H_k) \left( \frac{4}{9} a - 3 b - \frac{2}{15} c \right) \, .
\end{eqnarray}
 For smooth spatial boundary deformations $\Sigma(\epsilon)$ the boundary metric topology is unchanged and being toroidal $\chi(\Sigma)$ will vanish. 
Since the ODEs are linear in $A,B,C$, we may take $a = 1$ without loss of generality. 
We do not wish to make any assumption about how many components of the bulk horizon there are, or their topology, so we choose the constant $b$ as $b = \frac{2}{9} \left( \frac{2}{3} a  - \frac{1}{5} c\right)$, so that the term involving the Euler characteristic of the horizon components, $\chi(H_k)$, in the above equation has vanishing coefficient.
Finally we define the constant $k$,
\begin{eqnarray}
 k = \frac{2}{3} - \frac{c}{5} 
 \end{eqnarray}
which encodes the remaining constant of integration $c$, and 
then the sum of boundary terms becomes,
\begin{eqnarray}
\label{eq:Fboundsurface}
\int_{\partial \mathcal{M}} dA^a J_a & = &  \frac{2}{3 c_T} \left( E - k \, T S \right)  - \left( \frac{4}{3} - 2 k \right) \left( \frac{4 \pi T}{3}\right)^3 A(\Sigma)
\end{eqnarray}
where $S$ is the total entropy from all the horizon components.
We regard $k$ as fixing the energetic quantity of interest -- $k = 0$ giving energy, and $k = 1$ giving free energy. Then the question is whether we have $A(Z), C(Z) \ge 0$ for all $Z \ge 0$. 

Near $\partial \mathcal{M}_\infty$ one finds, $C(Z) = \frac{4 k}{3 Z^2} + \left( \frac{10}{3} - 5 k \right) + O(Z^1)$, and hence $k \ge 0$ is necessary for positive $C$. However, near the horizon one finds,
$C(Z) = \frac{2}{3 Z^3} \left( 2 - 5 k  \right) + O( Z^{-2})$
and so positivity of $C$ also requires $k \le \frac{2}{5}$.
This has the unfortunate consequence that we cannot find a bound in the case of the natural thermodynamic quantity, the free energy. 
However, for all $0 \le k \le \frac{2}{5}$ in fact the functions $A(Z)$, $C(Z) \ge 0$ for all $Z \ge 0$. Explicitly we may write these as,
\begin{eqnarray}
A(Z(\mu)) &=&   \frac{4 (1 - \mu) }{(2+\mu^3)^2} \left[ \left( 1 - \frac{k}{2} \mu^3 \right) + \left(1 - 2 k \right) \mu \left( 1 + \mu \right)  \right] \; , \nonumber \\
C(Z(\mu))  &=&  \frac{2}{3 z_0^3} \frac{(1-\mu^3)^{3/2}}{\mu^2 (2+\mu^3)^2}  \left[ 5 k \left( 1 - \mu^2 \right) + 10 \left( 1 - \frac{5 k}{2} \right) \mu^2 + 8  \left( 1 - \frac{5 k}{2} \mu \right) \mu^2 + 3 k \left( 1 - \mu^5 \right) + 2 \mu^2 \left( 1- \mu^3 \right)   \right] \nonumber \\
\end{eqnarray}
and over the domain $Z \ge 0$, so  $0 \le \mu \le 1$, we see each of the terms in the square brackets above is indeed positive provided $k$ is in the range $0 \le k \le \frac{2}{5}$.
Then for these $k$ we have $A, C \ge 0$,  and the bound~\eqref{eq:RobinsonBound} applies and from~\eqref{eq:Fboundsurface} yields,
\begin{eqnarray}
\label{eq:partialbound}
   \frac{1}{c_T}  \mathcal{F}_k(\epsilon)  - \left(2 - 3 k\right) \left( \frac{4 \pi T}{3}\right)^3 A(\Sigma(\epsilon))  \le 0 \, .
\end{eqnarray}
Now for two key points in the argument:
\begin{itemize} 
\item By construction this bound is saturated for the toroidal AdS-Schwarzschild solution with temperature $T$ and boundary $\Sigma(0)$.
\item Our family of spatial boundary metrics have fixed area $A(\Sigma(\epsilon)) = A(\Sigma(0))$.
\end{itemize}
The first point implies $ \frac{1}{c_T}  \mathcal{F}_k(0) = \left(2 - 3 k\right) \left( \frac{4 \pi T}{3}\right)^3 A(\Sigma(0))$, recalling that we chose boundary conditions that exclude the AdS-soliton so that the only solution for the flat boundary torus is AdS-Schwarzschild. Then combining it with the second implies,
\begin{eqnarray}
0 \ge \frac{1}{c_T}  \mathcal{F}_k(\epsilon)  - \left(2 - 3 k\right) \left( \frac{4 \pi T}{3}\right)^3 A(\Sigma(\epsilon)) = \frac{1}{c_T} \left( \mathcal{F}_k(\epsilon) - \mathcal{F}_k((0) \right) =  \frac{1}{c_T} \Delta \mathcal{F}_k(\epsilon)
\end{eqnarray}
and we obtain our  result stated earlier in  equation~\eqref{eq:result}. 
The main assumption underlying this bound is that some smooth static bulk solution exists with the prescribed spatial boundary geometry $\Sigma(\epsilon)$. 
We emphasise that no assumption concerning the number, or topology of bulk horizon components has been made, only that they are smooth Killing horizons with a common surface gravity determined by the temperature.

The fact that the bound is independent of the nature of bulk horizons follows from our choice of $b$ above.
We note that if instead of that choice, we  take $b = 0$, then one finds $A(Z), C(Z) \ge 0$ for all $Z \ge 0$ over the larger range $0 \le k \le \frac{6}{11}$ (which still does not include the $k=1$ case of free energy). Assuming either no horizon components, or a single connected horizon component with toroidal topology (as for AdS-Schwarzschild) then the boundary term involving $\chi(H_k)$ in equation~\eqref{eq:horizonterms} also vanishes, now either as there is no horizon or because its Euler characteristic vanishes. The same bound, $\frac{1}{c_T} \Delta\mathcal{F}_k(\epsilon) \le 0$, then is found for this greater range of $k$ but with the caveat that it makes an assumption about the nature of bulk horizons. 


\section{Summary and discussion}
\label{sec:discussion}

For a flat torus boundary metric a $4$-dimensional bulk solution to gravity plus negative cosmological constant is toroidal AdS-Schwarzschild. We have shown that 
if we deform the spatial boundary metric away from being flat, preserving staticity, and keeping temperature and spatial area fixed, then assuming a static bulk solution exists with the correct conformal boundary, its energy is less than that of the original undeformed toroidal AdS-Schwarzschild. In fact its `generalised energy', $\mathcal{F}_k =  E - k \, T S$, for $0 \le k \le 2/5$ behaves similarly, although this result does not extend to the free energy case $k=1$.

Let us reconsider the holographic interpretation of this result. The pure gravity plus cosmological constant is the `universal sector' of holographic CFTs (described by usual two-derivative bulk gravity theories). 
At finite temperature with a flat torus boundary, periodic fermion boundary conditions about the torus cycles, and no other sources for operators turned on, the CFT equilibrium thermal state is dual to toroidal AdS-Schwarzschild.  Hence we may translate the gravitational bound to the holographic statement that for a $3$-dimensional holographic CFT on a deformed spatial torus with periodic fermion boundary conditions and no other marginal or relevant deformations turned on, the equilibrium thermal state has generalised energy (for $0 \le k \le 2/5$)  less than that of the CFT at the same temperature on a flat torus of the same area.
Taking the area of the torus to be very large, while holding the temperature and scale of the metric deformation fixed, this result indicates the (generalised) energy for deformations of infinite flat space is less than that of flat space, where the fixed area condition in infinite volume becomes the requirement that the area functional is invariant under the deformation, ie. $\int_\Sigma d^2x \frac{d}{d\epsilon} \sqrt{\bar{g}(\epsilon)} = 0$.

This result extends the explicit calculations of~\cite{Cheamsawat:2018wkr} for the quadratic variation of free energy (and energy as discussed here) of $3$-dimensional holographic CFTs.  Our results here show that the decrease in energy seen perturbatively in that work is in fact non-perturbative in the metric for $3$-dimensional holographic CFTs -- flat space globally maximises energy at fixed temperature when preserving area. This would be a remarkable statement to be able to make in field theory -- it is difficult to imagine making such a claim even in free field theory, where understanding the perturbative quadratic variation of energy is straightforward, but moving to higher perturbative orders or full non-perturbative results becomes very challenging technically. This is an example of where very powerful results in curved spacetime field theory can be derived in the holographic context~\cite{Marolf:2013ioa}.

We have until now assumed periodic boundary conditions for bulk fermion fields so that for the flat toroidal boundary the AdS-soliton is not an admissible solution. This has simplified the discussion, but is not strictly necessary. Suppose now we relax the requirement on the boundary conditions. Then at sufficiently low temperature the AdS-soliton may dominate the bulk partition function as a saddle point, having lower free energy than that of the toroidal AdS-Schwarzschild. This solution has spacetime metric,
\begin{eqnarray}
\label{eq:soliton}
ds^2_{(4)} = \frac{\ell^2}{z^2} \left( - dt^2 + dx^2  + \left( 1 - \left( \frac{z}{z_c} \right)^3 \right) dy^2 + \left( 1 - \left( \frac{z}{z_c} \right)^3 \right)^{-1}  dz^2 \right)
\end{eqnarray}
where the bulk has no horizons, and smoothly `caps' off as the torus cycle generated by $\partial / \partial y$ degenerates at $z = z_c$, and $z_c$ is related to the cycle period as $L = \frac{4}{3} \pi z_c$. For this solution $Z = z$ and the optical Ricci scalar is $R(Z) = 6 Z/z_c^3$.
This solution has negative energy density,
\begin{eqnarray}
\frac{1}{c_T} \rho = - \frac{1}{z_c^3} \, .
\end{eqnarray}
and at sufficiently low temperature, so that $z_c < z_0$, the AdS-soliton (whose energy is equal to its free energy as it has no horizon and hence zero entropy) dominates the bulk partition function. In this case we have to be more careful about the meaning of $\mathcal{F}_k(0)$, as while the toroidal AdS-Schwarzschild solution exists, really $\mathcal{F}_k(0)$ should be given by the generalised energy of the AdS-soliton and then the arguments leading to the bound, as stated above, do not apply.
This requires only a simple modification of the statement of the bound.
Firstly we observe that the bound in~\eqref{eq:partialbound} does not assume anything about the number or topology of bulk horizon components, and therefore also applies in the case of the AdS-soliton or its deformations as the boundary torus is deformed from being flat. Thus we may obtain a similar bound $\frac{1}{c_T} \Delta'\mathcal{F}_k \le 0$ for $0 \le k \le \frac{2}{5}$ where now $\Delta'\mathcal{F}_k = \mathcal{F}_k(\epsilon) - \mathcal{F}^{(\mathrm{Sch})}_k$, and $\mathcal{F}^{(\mathrm{Sch})}_k$ is the generalised energy of AdS-Schwarzschild at the appropriate temperature and with flat boundary torus with the appropriate area.
While the context for this work is understanding how the thermal equilibrium energy with a deformed boundary space compares to that with a flat boundary, this bound is also interesting even restricting to a flat boundary torus. For example, it states that the energy of a toroidally compactified localised `plasmaball'~\cite{Aharony:2005bm,Figueras:2014lka} (i.e. an AdS-soliton containing a `small' black hole) cannot exceed that of the AdS-Schwarschild solution at the same temperature.
Finally, we note that in the holographic context we may naturally state the bound as saying that at a given temperature the generalised free energy of the CFT on a deformed torus with prescribed fermion boundary conditions is less than that of the CFT on a flat torus \emph{with periodic fermion} boundary conditions and the same area.

One might  wonder if in the case  of temperature and area where the preferred flat torus dual is the AdS-soliton one could prove the energy or generalised energy is lower when the boundary metric is deformed. This is different to our result which just shows the energy or generalised energy is lower than that of the toroidal AdS-Schwarzschild (which in this situation would be true already for the flat torus with dual given by the AdS-soliton). Using the relation $R(Z)$ for the soliton one can define a suitable function $H(Z)$ so that the function $\Phi(x)$ in equation~\eqref{eq:phi} vanishes for the AdS-soliton. However we then have not found a solution to the resulting ODE system that has positive $A(Z)$ and $C(Z)$. It would be interesting to understand how the energy of the soliton varies as its boundary metric is deformed.

An important concern is that for substantial metric deformations perhaps a new bulk black hole solution exists with lower free energy, outside the universal sector (ie. involving condensation of some other bulk fields, or non-trivial configurations on an internal space), and it is this that dominates the partition function. Had the bound we found been for the free energy, rather than the energy, then we could be confident that even if this occurred, the free energy of the true vacuum would still obey our bound, since if for the thermal state in the universal sector the free energy is less than that for flat torus boundary, then it would have to be true for the true vacuum too, as to dominate it would have to have an even lower free energy. However this argument cannot be made in our case for the energy (or the generalised energies $\mathcal{F}_k$) since it is in principle possible to have a true vacuum bulk solution with lower free energy but higher energy than the thermal vacuum state in the universal sector. 

Clearly it would be good to be able to prove a bound for the free energy rather than the energy (or range of generalised energies we have here). Given the relatively complicated construction of the proof, and since we are able to obtain bounds for the generalised energies up to $k = 2/5$ we feel it is likely that by modifying the proof an improvement in the range of $k$ could be achieved. One possible direction would be a more general version of the starting point in equation~\eqref{eq:J} as recently explored in the asymptotically flat setting~\cite{Nozawa:2018kfk}. 
The intuition from free field theory suggests that a gravitational upper bound could be found for the thermal equilibrium free energy, at least in the infinite volume limit.
However there is an important caveat. Our bound states that the generalised energy of a solution with deformed boundary torus is not greater than that of AdS-Schwarzschild for the same boundary area and temperature. However, for free energy this obviously fails even for the flat torus at high temperatures (relative to the torus size) as the soliton exists as a solution but precisely fails to be the dominant saddle point in the partition function as it has greater free energy than that of AdS-Schwarzschild.
Considering the generalised energy $\mathcal{F}_k = E - k \,T S$, one finds that for $k \ge \frac{2}{3}$ the soliton has a larger value than that of AdS-Schwarzschild at sufficient temperature. For such values of $k$ then clearly we cannot have a bound stating the generalised energy of \emph{any} bulk solution is always less than that of AdS-Schwarzschild. 
Rather we might hope to prove there exists \emph{some} bulk solution with generalised energy less than that of AdS-Schwarzschild. 
Thus in order to extend beyond $k = \frac{2}{3}$, towards the case of free energy, an essential part of the argument would have to be changed.
Perhaps a bound might hold for solutions which necessarily have some bulk horizon, or more specifically a toroidal horizon. Alternatively a bound might only apply for infinite volume.

The bound we have reached is of a rather novel type for black holes. In asymptotically flat space static black holes are rigid in the sense that there is typically a unique solution. In the case of AdS asymptotics, the freedom to deform the boundary geometry leads to a much richer structure of static black hole solutions and we can regard our bound as a restriction on the behaviour of such families. It is worth emphasising that it is quite different in spirit to energy bounds that apply to different interior solutions with a \emph{fixed} conformal boundary geometry (such as, for example \cite{Chrusciel:2000az,Galloway:2015ora} where the boundary spatial geometry has constant curvature).

Finally we emphasise an important point in this work has been the observation that it is natural to fix both temperature and spatial boundary area as we deform the boundary metric, and to focus on the case of a 3-dimensional boundary.
For these observations the intuition gained from the behaviour of free field theory at finite temperature on deformed spaces was crucial. It is fascinating that, via AdS-CFT, we may discover entirely new results in black holes physics starting from free field theory considerations. 
\\

\noindent
{\it Note added:}
After the appearance of the preprint version of this paper on the arXiv, there appeared
\cite{Chrusciel:2019lgm} which observed that for a static black hole solution to the gravitational equations we consider here, with spatial conformal boundary a torus (or indeed any compact geometry), as we are interested in, then the free energy of the solution is negative in the case that the horizon has torus or higher genus topology. In fact this result also  follows from equation~\eqref{eq:bound1} in our review section since for a horizon of higher genus than a sphere then $\chi(H) \le 0$. It is found in the earlier~\cite{HicWis15} for general compact spatial conformal boundary (and~\cite{Galloway:2015ora} in the special case of constant curvature spatial boundaries, ie. a flat torus, round sphere etc...).
However this result in itself gives no information of how the free energy \emph{changes} when a flat torus boundary metric
is deformed, which is the focus of this paper. Our main results concern this change, $\Delta \mathcal{F}_k$, not the quantity $\mathcal{F}_k$ itself.
We thank a referee for pointing out the need to clarify the relation between the work of the present paper and the subsequent results of \cite{Chrusciel:2019lgm}.


\section*{Acknowledgements}

We would like to thank the Fields institute, Toronto, where this work was initiated during the `Workshop on General Relativity and AdS/CFT'.
TW is supported by the STFC grant ST/P000762/1. KC is sponsored by the DPST scholarship from the Royal Thai Government.


\bibliographystyle{apsrev4-1}
\bibliography{RobinsonBib}

\end{document}